# A one-dimensional Lattice Boltzmann method for modeling the dynamic pole-to-pole oscillations of Min proteins for determining the position of the midcell division plane

Waipot Ngamsaad, Wannapong Triampo[*], Paisan Kanthang, I-Ming Tang,
Narin Nuttawut and Charin Modjung

*Department of Physics and Capability Building Unit in Nanoscience and Nanotechnology, Faculty of Science, Mahidol University, Bangkok 10400, Thailand*

Yongwimon Lenbury

*Department of Mathematics, Faculty of Science, Mahidol University, Bangkok 10400, Thailand*

**Abstract**

Determining the middle of the bacteria cell and the proper placement of the septum is essential to the division of the bacterial cell. In *E. coli,* this process depends on the proteins MinC, MinD, and MinE. Here, the Lattice Boltzmann method (LBM) is used to study the dynamics of the oscillations of the *min* proteins from pole to pole. This determines the midcell division plane at the cellular level. The LBM is applied to the set of the deterministic reaction diffusion equations proposed by Howard *et. al.* [1] to describe the dynamics of the Min proteins. The LBM results are in good agreement with those of Howard *et al,* and agree qualitatively with the experimental results. Our good results indicate that the LBM can be an alternative computational tool for simulating problems dealing with complex biological system which are described by the reaction-diffusion equations.

PACS numbers: 87.15.Aa

Keywords: Lattice Boltzmann method, bacteria, *E. coli,* cell division, *min* proteins, MinCDE oscillation

*Electronic address: wtriampo@yahoo.com; Fax: 662-201-5843

# I. Introduction

Cell division or cytokinesis is the process by which a cell separates into two after its DNA has been duplicated and distributed into the two regions, which will become the future daughter cells. For a successful cell division to take place, the cell has to determine the optimal location of the cell separation and the time to start the cell cleavage. This involves the identification of the midpoint of the cell where the septum or cleavage furrow will form. For *Escherichia coli* and other rod-like bacteria, evidence accumulated over the past few years indicate that the separation into two daughter cells is achieved by forming a septum perpendicular to its long axis. To induce the separation, the FtsZ ring (Z ring), a tubulin-like GTPase is believed to initiate and guide the septa growth by contraction [2]. The Z ring is usually positioned close to the center, but it can also form in the vicinity of the cell poles. Two processes are known to regulate the placement of the division site: nucleoid occlusion [3] and the action of the *min* proteins [4]. Both processes interfere with the formation of the Z ring, which is believed to determine the division site. Nucleoid occlusion is based on cytological evidence that indicates that the Z ring assembles preferentially on those portions of the membrane that do not directly surround the dense nucleoid mass [5].

The *min* proteins that control the placement of the division site are the MinC, MinD, and MinE proteins [4]. Experiments, involving the use of modified proteins show that MinC is able to inhibit the formation of the FtsZ-ring [6]. MinD is an ATPase that is connected peripherally with the cytoplasmic membrane. It can bind to MinC and activate the function of the MinC [7,8]. Recent studies show that MinD recruits MinC to the membrane. This suggests that MinD stimulates the MinC by concentrating it near its presumed site of activation [9,10]. MinE is required to give site specificity to the division inhibitor suggests that MinE acts as a topological specificity protein, capable of recognizing the midcell site and preventing the MinC division inhibitor from acting at this site [11]. Its expression results in a site-specific suppression of the MinC/MinD action so that the FtsZ assembly is allowed at the middle of the cell but is blocked at other sites [4]. In the absence of MinE, the MinC/MinD is distributed homogeneously over the entire membrane. This results in a complete blockage of the Z-ring formation and the subsequent formation of the long filamentous cell which would fail to divide [9,10,12,13]. By fluorescent labeling, MinE was shown to attach to the cell wall only in the presence of MinD [14,15]. Because MinD interacts with MinC, it is likely that they oscillate together. This results in the concentration of the division inhibitor at the membrane on either cell end alternating between being high or low every other 20 second, so that the period of oscillation is about 40 second per cycle [9,10]. MinE is not only required for the MinC/MinD oscillation, it is also involved in setting the frequency of the oscillation cycle [12]. Several lines of evidence indicate that the MinE localization cycle is tightly coupled to the oscillation cycle of MinD. Recently, microscopy of fluorescently labeled proteins involved in the regulation of *E. coli* division have uncovered coherent and stable spatial and temporal oscillations of these three proteins [16]. The proteins oscillate from end to end of the bacterium, and move between the cytoplasmic membrane and cytoplasm. The detail mechanism by which these proteins determine the correct position of the division plane is currently unknown, but the observed pole-to-pole oscillations of the corresponding distribution are thought to be of functional importance.

## II. Lattice Boltzmann method and model description

The Lattice Boltzmann method (LBM) is a numerical scheme evolved from the lattice gas model in order to overcome the difficulties encountered with the lattice gas model (LGM) [17,18]. LGM or lattice gas automata are a method to determine the kinetics of particles utilizing a discrete lattice and discrete time. It has provided insights into the underling microscopic dynamics of the physical system, whereas most other approaches focus only on the solution of the macroscopic equation. However, LGM in which the particles obey an exclusion principle has microscopic collision rules. These rules are very complicate and require many random numbers. These random numbers create noise or fluctuations. An ensemble averaging is then required to smooth out the noise in order to obtain the macroscopic dynamics which are the results of the collective behavior of the many microscopic particles in the system and which are not sensitive to the underlying details at the microscopic level. The averaging requires long time averages that lead to an increase in the amount of computational storage required and which in turn leads to a reduction in the computational speed. For these reasons, the LBM is used whenever one is only interested in the evolution of averaged quantities and not in the influence of the fluctuations. LBM gives a correct average description on the macroscopic level of a fluid. The LBM can also be viewed as a special finite difference scheme for kinetic equation of the discrete-velocity distribution function. The simplicity and kinetic nature of the LBM are among its appealing features.

LBM consists of simple arithmetic calculations and is therefore easy to program. In LBM, the space is divided into a regular Cartesian lattice grid as a consequence of the symmetry of the discrete velocity set. Each lattice point has an assigned set of velocity vectors with specified magnitudes and directions connecting the lattice point to its neighboring lattice points. The total velocity and particle density are defined by specifying the number of particles associated with each of the velocity vectors. The microscopic particle distribution function which is the only unknown evolves at each time step through a two-step procedure: convection and collision. The first step, convection (or streaming process), is to simply advance the particles from one lattice site to another lattice site along the directions of motion according to their velocities. This feature is borrowed from kinetic theory. The second step or collision is to model various interactions among particles by allowing for the relaxation of a distribution towards an equilibrium distribution using a linear relaxation parameter. The averaging process uses information based on the whole velocity phase space.

Most research reported in the literature are limited to the LBM for the Navier-Stokes equations [19,20]. The LBM scheme has been particularly successful in simulating fluid flow applications for a broad variety of complex physical systems and has found application in different areas, such as hydrodynamic systems [18,22], multiphase and muti-component fluids [21], advection-dispersion [23] and blood flow [24,25,26]. Application to complex biological systems at the cellular and the molecular biological levels has been rare.

In the present paper, we propose a LBM to study the partitioning of the bacterial cell during cell division. This provides an alternative method to investigate quantitatively the division of the cell. We compare our results with those obtained by numerically solving a set of deterministic coarse-grained coupled reaction diffusion equations [1] to demonstrate the validity of the proposed LBM.

**II.a    Reaction-diffusion equation model**

We focus on the *E. coli* bacteria, a commonly studied rod shaped bacteria of approximately $2-6\,\mu m$ in length and around $1-1.5\,\mu m$ in diameter. Each *E. coli* bacteria divides roughly every hour via cytokinesis. We adopted the dynamic model of the compartmentization in the bacterial cell division process proposed by Howard *et. al*. In the Howard model, dynamics at the mean-field level are given by a set of coarse-grained non-linear reaction-diffusion equations. The reaction-diffusion equations have often been used in biological applications to model self-organization and pattern formation [27].

Our starting point is the set of one dimensional deterministic coupled reaction-diffusion equations used to describe the dynamics of the interactions between the local densities of MinD and MinE proteins given by Howard *et al* [1]. They describe the time rates of change of the densities due to the diffusion of the MinD and MinE and to the mass transfer between the cell membrane and the cytoplasm. Based on the experimental results given in [10], which showed that the MinC dynamics are similar to those of MinD, we have not written out the equations for MinC. In dimensionless form, the dynamics are written as:

$$\frac{\partial n_D}{\partial t} - D_D \frac{\partial^2 n_D}{\partial x^2} = R_D = -\frac{\sigma_1 n_D}{1+\sigma'_1 n_e} + \sigma_2 n_e n_d \ , \qquad (1)$$

$$\frac{\partial n_d}{\partial t} - D_d \frac{\partial^2 n_d}{\partial x^2} = -R_D = \frac{\sigma_1 n_D}{1+\sigma'_1 n_e} - \sigma_2 n_e n_d \ , \qquad (2)$$

$$\frac{\partial n_E}{\partial t} - D_E \frac{\partial^2 n_E}{\partial x^2} = R_E = \frac{\sigma_4 n_e}{1+\sigma'_4 n_D} - \sigma_3 n_D n_E \ , \qquad (3)$$

$$\frac{\partial n_e}{\partial t} - D_e \frac{\partial^2 n_e}{\partial x^2} = -R_E = -\frac{\sigma_4 n_e}{1+\sigma'_4 n_D} + \sigma_3 n_D n_E \qquad (4)$$

where $n_s$ is the mass density of particle of species $s = \{D, d, E, e\}$ at time $t$ and position $x$. The first equation is for the cytoplasmic MinD density $n_D$. The second is for the membrane bound MinD density $n_d$. The third is for the cytoplasmic MinE density $n_E$ and the last is for the membrane bound MinE density $n_e$. $R_s$ is the reaction term which depends on the density of the species $n_s$ and on the densities of the other species that react with species $s$. $D_s$ is the diffusion coefficient In this paper, we assume that $D_s$ is isotropic and independent of $x$. The constant $\sigma_1$ represents the association of MinD to the membrane [13]. $\sigma'_1$ corresponds to the membrane-bound MinE suppressing the recruitment of MinD from the cytoplasm. $\sigma_2$ reflects the rate that MinE on the membrane drives the MinD on the membrane into the cytoplasm. Based on the evidence of the cytoplasmic interaction between MinD and MinE [8], we let $\sigma_3$ be the rate that cytoplasmic MinD recruits the cytoplasmic MinE to the membrane while $\sigma_4$ corresponds to the rate of dissociation of MinE from the membrane to the cytoplasm. Finally, $\sigma'_4$ corresponds to the cytoplasmic MinD suppressing the release of the membrane-bound MinE. The time scale of the diffusion on the membrane is much slower than that in cytoplasm. It seems therefore reasonable to set $D_d$ and $D_e$ to zero. In this dynamics, we allow for the Min protein to bind/unbind from the membrane but not to be degraded in the process. Thus the total amount of each type of Min proteins is conserved. The zero flux boundary condition will be imposed. This boundary condition gives a closed system with reflecting or hard-wall boundary conditions.

### II.b  Lattice Boltzmann equation

The dynamics determined by eqns. (1)-(4) can be simulated using a Lattice-Boltzmann method having three one dimensional velocities. Let $f_s(\bar{x}, i, t)$ be the one-particle distribution function of species $s$ with velocity $\bar{e}_i$ at some dimensionless time, $t$ and dimensionless position $\vec{x}$. The coordinate $\vec{x}$ only takes on a discrete value: the nodes of the chosen lattice. The nearest neighbor vectors are defined as

$$\vec{e}_i = \begin{cases} \vec{0} & i = 0, \\ \hat{x}, & i = 1, \\ -\hat{x}, & i = 2, \end{cases} \tag{5}$$

where $\hat{x}$ is an unit vector along the $x$ direction. For each lattice site, we have three states for each species. Following [28], the lattice Boltzmann equation for $f_s(\vec{x}, i, t)$ can be written as

$$f_s(\vec{x} + \vec{e}_i, i, t+1) - f_s(\vec{x}, i, t) = \Omega_s(\vec{x}, i, t) \tag{6}$$

where $\Omega_s$ is the collision operator for the species $s$ and depends on the distribution function $f_s$. The collision operator $\Omega_s$ can be separated into two parts [29], a nonreactive term ($\Omega_s^{NR}$) and a reactive term ($\Omega_s^R$), i.e.,

$$\Omega_s = \Omega_s^{NR} + \Omega_s^R. \tag{7}$$

In order to relate the results obtained by solving Eq. (6) with the solutions of eqns. (1)-(4), we need to derive the evolution equations for the moments of the function, $f_s$. The zeroth moment of $f_s$, the total number of particles of species $s$ at time $t$ and position $x$, is defined as

$$n_s(\vec{x}, t) \equiv \sum_i f_s(\vec{x}, i, t) = \sum_i f_s^{eq}(\vec{x}, i, t). \tag{8}$$

For the nonreactive term, $\Omega_s^{NR}$ we use the BGK approximation with a single relaxation time $\tau_s$ [30]

$$\Omega_s^{NR} = -\frac{1}{\tau_s}\left[f_s(\vec{x}, i, t) - f_s^{eq}(\vec{x}, i, t)\right], \tag{9}$$

where the equilibrium distribution function of species $f_s^{eq}(\vec{x}, i, t)$ depends on $\vec{x}$ and $t$ through the local density and velocity. Here we use the simple equilibrium distribution function corresponding to a system with zero mean flow as follow:

$$f_s^{eq} = w_{s,i} n_s, \tag{10}$$

where the weights $w_{s,i}$ is dependent on the lattice symmetry [31]. We can write

$$w_{s,i} = \begin{cases} z_s & i = 0 \\ (1 - z_s)/2 & i = 1, 2 \end{cases}, \tag{11}$$

where $z_s$ denotes the fraction of particles at rest and which can be different for the different species. For the reactive term $\Omega_s^R$, we use the simple isotropic form [31]

$$\Omega_s^R = w_{s,i} R_s \qquad (12)$$

where $R_s$ is non-linear reaction term and depends on the densities of the reacting species. Thus, it couples the Boltzmann equations for the different species. The choice given in Eq. (12) is the simplest choice that can provide the right macroscopic solution using LBM (as we shall see later).

To show that lattice Boltzmann equation is valid for reacting system, we employ a procedure called the Chapmann-Enskog expansion [18]. We, first, expand the left hand side of Eq. 6 via a Taylor series

$$f_s(\vec{x}+\vec{e},i,t+1) - f_s(\vec{x},i,t) \cong \frac{\partial f_s(\vec{x},i,t)}{\partial t} + e_i \frac{\partial f_s(\vec{x},i,t)}{\partial x} + \frac{1}{2} e_i^2 \frac{\partial^2 f_s(\vec{x},i,t)}{\partial x^2} = \Omega_s . \qquad (13)$$

We then expand $f_s$ about the equilibrium distribution function in terms of the parameter $\varepsilon$

$$f_s \cong f_s^{eq} + \varepsilon f_s^{(1)} \qquad (14)$$

We now assume [29]:

$$\frac{\partial}{\partial x} \to \varepsilon \frac{\partial}{\partial x} \qquad (15)$$

$$\frac{\partial}{\partial t} \to \varepsilon^2 \frac{\partial}{\partial t} \qquad (16)$$

$$R_s \to \varepsilon^2 R_s . \qquad (17)$$

Substituting eqns. (15), (16) and (17) into eqn. (13), we obtain

$$e_i \frac{\partial f_s^{eq}(\vec{x},i,t)}{\partial x} = -\frac{f_s^{(1)}(\vec{x},i,t)}{\tau_s} \qquad (18)$$

to order $\varepsilon^1$ and

$$\frac{\partial f_s^{eq}(\vec{x},i,t)}{\partial t} + e_i \frac{\partial f_s^{(1)}(\vec{x},i,t)}{\partial x} + \frac{1}{2} e_i^2 \frac{\partial^2 f_s^{eq}(\vec{x},i,t)}{\partial x^2} = w_{s,i} R_s \qquad (19)$$

to order $\varepsilon^2$. From eqn. (18), we immediately obtain

$$f_s^{(1)}(\vec{x},i,t) = -\tau_s w_{s,i} e_i \frac{\partial n_s}{\partial x} \qquad (20)$$

Inserting eqn. (20) to eqn. (19) and doing some simple algebra, we have, to order $\varepsilon^2$,

$$\frac{\partial n_s}{\partial t} - \left(\tau_s - \frac{1}{2}\right) e_i^2 \frac{\partial^2 n_s}{\partial x^2} = R_s . \qquad (21)$$

Eliminating the $e_i^2$ term by carrying out an averaging with weight $w_{s,i}$, we get

$$\frac{\partial n_s}{\partial t} - (1-z_s)\left(\tau_s - \frac{1}{2}\right)\frac{\partial^2 n_s}{\partial x^2} = R_s, \qquad (22)$$

which is the dimensionless version of the initial reaction-diffusion equation.

To summarize, we will now implement the numerical evaluation in two steps

- *Collision step*: $\tilde{f}_s(\vec{x},i,t+1) = f_s(\vec{x},i,t) - \frac{1}{\tau_s}\left[f_s - f_s^{eq}\right] + w_{s,i} R_s$

- *Streaming step*: $f_s(\vec{x}+\vec{e}_i,i,t+1) = \tilde{f}_s(\vec{x},i,t+1)$.

The boundary treatment is an important issue in the LBM simulation and advancement are still being made [32,33]. Here we use the impermeable boundary suggested by Zhang *et al.*, [34].

## III. Numerical results and discussion

To demonstrate the validity of the proposed LBM applied to the Howard dynamic model for determining the partition of *E. coli* mediated by *min* proteins, we implemented the LBM as given in the previous section on a PC using C programming. In the simulation, we use the same parameters given by Howard *et al.* The 2 micron long bacterium is divided into 250 grids. The discrete space steps are therefore $dx = 0.008\,\mu m$. Time step of $dt = 6.4\times10^{-5}$ s are chosen. The dimensionless parameters are $D_D = 0.28$, $D_E = 0.6$, $D_d = D_e = 0$, $\sigma_1 = 1.28\times10^{-3}$, $\sigma_4 = 5.12\times10^{-5}$, $\sigma_2 = 4.032\times10^{-7}$, $\sigma_3 = 2.56\times10^{-6}$, $\sigma'_1 = 0.028$, $\sigma'_4 = 0.027$. The relaxation time is calculated by eqn. (22) and is given as $\tau_s = D_s/(1-z_s) + 0.5$. The initial number of MinD and MinE is randomly initialized as 3000 for $n_D$ and 170 for $n_E$. Each simulation takes 156,250,000 iterations for $10^4$ s of the time division of the bacterium. We test the system with two possible sets of rest particle fraction $z_s = 1/3$ and 2/3 for all species. We found that $z_s = 2/3$ gives the more accurate result. We now present some results to show the validity and accuracy of our LBM and compare them with the results obtained from the deterministic reaction-diffusion equations approach.

In Fig. 1, the space-time plots of the MinD and MinE concentrations for a cell of length $2\,\mu m$ are shown. They are in qualitative agreement with the simulation obtained by Howard *et al.*, and are also in agreement with the experimental results. The MinE forms a line up in the middle of the cell and then sweeps towards a cell pole, displacing the MinD, which then reforms at the opposite pole. In Fig. 2, we plot the time averaged MinD and MinE densities as a function of position. These are again in excellent agreement with those given in by Howard *et al.* The results in both works are also in excellent agreement with the experimental data of Hale *et al.* The MinE concentration peaks at the mid cell and has minimum at the cell rims, with MinD virtually out of phase with MinE.

## IV. Concluding remarks

In this paper, we have proposed a new LBM approach to investigate the dynamic pole-to-pole oscillations of *min* proteins used to determine the middle of the bacterial cell division. We have developed a numerical scheme based on the LBM to simulate the coarse-grained coupled reaction-diffusion equations model used to describe the MinD/MinE interaction. It is found that our results are in good agreement with those given by Howard *et al*. The results are also in qualitative agreement with experimental results, in particular the oscillatory pattern of *min* proteins [35].

The LBM approach provides an alternative fast computational tool to study the protein oscillation. We believe that the LBM is an useful scheme for simulating at the cellular lever those biological system which are governed by the reaction-diffusion equations. In a future work, we will generalize the current LBM so that it can be used to study the effects of the inhomogeniety in the intracellular space and the possibility of the asymmetrical cell division.

**Acknowledgements**

We thank M. Howard, J. Wong-ekkabut and M. Chooduang for their useful comments and suggestion. This research is supported in part by Thailand Research Fund through the grant number TRG4580090 and RTA4580005. The IRPUS Program 2547 to Charin Modjung and W. Triampo is acknowledged. The Development and Promotion of Science and Technology talents program to Waipot Ngamsaad.

**References**

[1] M. Howard, A. D. Rutenberg and S. de Vet, Phys. Rev. Lett. **87,** 278102 (2001).

[2] J. Lutkenhaus, Mol. Microbiol. **9,** 403 (1993).

[3] C. L. Woldringh, E. Mulder, P. G. Huls and N. Vischer, Res. Microbiol. **142,** 309 (1991).

[4] P. A. J. de Boer, R. E. Crossley and L. I. Rothfield, Cell. **56,** 641 (1989).

[5] E. Mulder and C. L. Woldingh, J. Bacteriol **171,** 4303 (1989).

[6] P. A. J. de Boer, R. E. Crossley and L. I. Rothfield, PNAS (USA) **87,** 1129 (1990).

[7] P. A. J. de Boer, R. E. Crossley, A. R. Hand and L. I. Rothfield, EMBO J. **10,** 4371 (1991).

[8] J. Huang, C. Cao and J. Lutkenhaus J. Bacteriol. **178,** 5080 (1996).

[9] Z. Hu and J. Lutkenhaus, Mol. Microbiol. **34,** 82 (1999).

[10] D. M. Raskin and P. A. J. de Boer, J. Bacteriol. **181,** 6419 (1999).

[11] X. Fu, Y. –L. Shih, Y. Zhang and L. I. Rothfield, PNAS(USA) **98**, 980 (2001).

[12] D. M. Raskin and P. A. J. de Boer, PNAS. (USA) **96,** 4971 (1999).

[13] S. L. Rowland, X. Fu, M. A. Sayed, Y. Zhang, W. R. Cook and L. I. Rothfield, J. Bacteriol. **182,** 613 (2000).

[14] K. C. Huang, Y. Meir, N. S. Wingreen, PNAS (USA) **100,** 12724 (2003).

[15] D. M. Raskin and P. A. J. de Boer, Cell. **91** 685 (1997).

[16] C. A. Hale, H. Meinhardt and P. A. J. de Boer, EMBO J. **20,** 1563 (2001).

[17] R. Benzi, S. Succi, M. Vergassola, Phys. Rep. **222**, 145 (1992).

[18] S. Chen, G. D. Doolen, Ann. Rev. Fluid Mech. **30,** 329 (1998).

[19] Y. H. Qian, D. d'Humieres and P. A. Lallemand, Europhys. Lett **17,** 479 (1992).

[20] H. Chen, S. Chen, and W. H. Matthaeus, Phys. Rev. A **45**, R5339 (1992)**.**

[21] N. S. Martys and H. D. Chen, Phys. Rev. E **53,** 743 (1996).

[22] G. D. Doolen, Lattice Gas Methods: Theory, Applications and Hardware, 2nd ed.; MIT: Cambridge, MA, (1991).

[23] R. G. M. van der Sman and M. H. Ernst, J. Comput. Phys**. 60,** 766 (2000).

[24] C. Migliorini, Y. H. Qian, H. Chen, E. Brown, R. Jainand and L. Munn, Biophys. J. **84,** 1834 (2002)**.**

[25] C. H. Sun, C. Migliorini and L. Munn, Biophys. J. **85,** 208 (2003).

[26] M. Hirabayashi, M. Ohta, D. A. Rufenacht and B. Chopard, FGCS **20,** 925 (2004).

[27] G. Nicolis and I. Prigogine, Self organization in Nonlinear Systems, Wiley, New York (1977).

[28] G. McNamara and G. Zanetti, Phys. Rev. Lett. **61,** 2332 (1988).

[29] S. P. Dawson, S. Chen and G. D. Doolen, J. Chem. Phys. **98**, 1514 (1993)**.**

[30] P. L. Bhatnagar, E. P. Gross and M. Krook, Phys. Rev. **94**, 511 (1954).

[31] R. Blaak, P. M. Sloot, Comp. Phys. Comm. **129** 256 (2000).

[32] S. Chen, D. O. Martinez and R. Mei, Phys Fluids **8**, 2527 (1996).

[33] Q. Zou and X. He, Phys Fluids **9,** 6202 (1997)**.**

[34] X. Zhang, J. W. Crawford, A. G. Bengough and I. M. Young, Adv. Water Resour. **25,** 601 (2002).

[35] H. Meinhardt and P. A. J. de Boer, Proc. Natl. Acad. Sci. **98**, 14202 (2001).

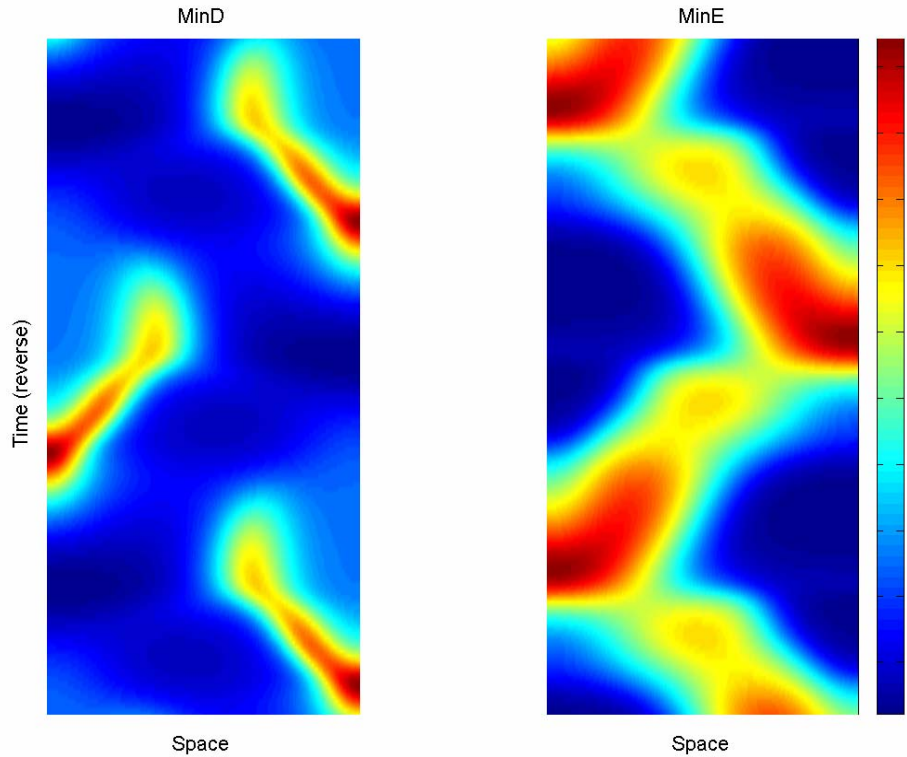

Fig. 1: Space-time plots of the total MinD (left) and MinE (right) densities. The color scale runs from the lowest (blue) to the highest (red). The MinD depletion from midcell and the MinE enhancement at the midcell are immediately evident. Times increase from top to bottom, and the pattern repeats indefinitely as time increases. The vertical scale spans the time 1000 second. The horizontal scale spans the bacterial length ( $2\,\mu m.$ )

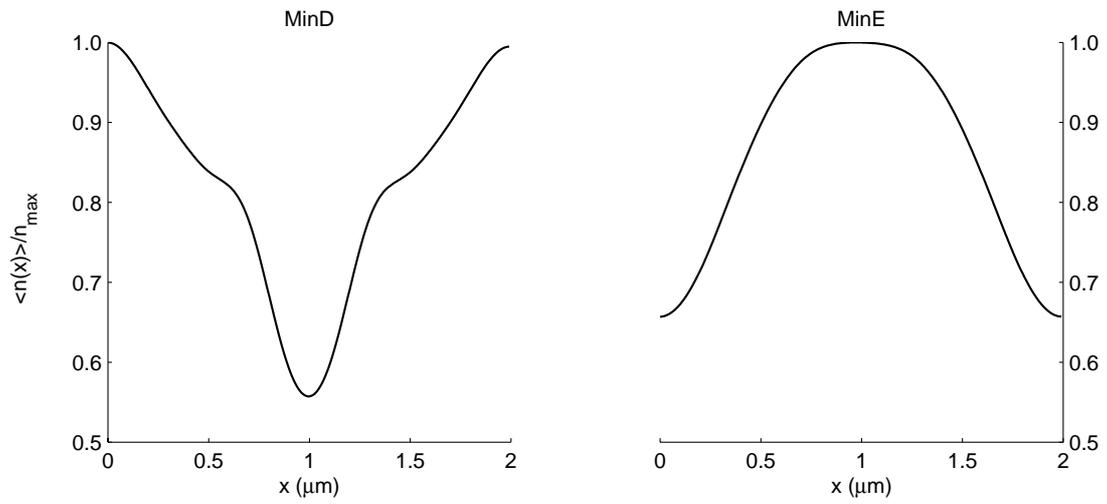

Fig. 2: The time average MinD(left) and MinE(right) densities $\langle n(x) \rangle / n_{max}$, relative to their respective time-average maxima, as a function of position $x$ (in $\mu m$) along the bacterium.